\documentstyle[prl,twocolumn,aps]{revtex}
\begin{document}
\draft

\title{H-Theorem and Generalized Entropies \\
Within the Framework of Non Linear Kinetics}
\author {G. Kaniadakis}
\address{Dipartimento di Fisica - Politecnico di Torino -
Corso Duca degli Abruzzi 24, 10129 Torino, Italy \\ Istituto
Nazionale di Fisica della Materia - Unit\'a del
 Politecnico di Torino. }

\date{\today}
\maketitle

\begin {abstract} In the present effort we consider the most general
non linear particle kinetics within the framework of the
Fokker-Planck picture. We show that the kinetics imposes the form
of the generalized entropy and subsequently  we demonstrate the
H-theorem. The particle statistical distribution is obtained, both
as stationary solution of the non linear evolution equation and as
the state which maximizes the generalized entropy. The present
approach allows to treat the statistical distributions already
known in the literature in a unifying scheme. As a working example
we consider the kinetics, constructed by using the
$\kappa$-exponential $\exp_{_{\{{\scriptstyle \kappa}\}}}(x)=
\left(\sqrt{1+\kappa^2x^2}+\kappa x\right)^{1/\kappa}$ recently
proposed which reduces to the standard exponential as the
deformation parameter $\kappa$ approaches to zero and presents the
relevant power law asymptotic behaviour $\exp_{_{\{{\scriptstyle
\kappa}\}}}(x){\atop\stackrel{\textstyle\sim}{\scriptstyle
x\rightarrow \pm \infty}}|2\kappa x|^{\pm 1/|\kappa|}$. The
$\kappa$-kinetics obeys the H-theorem and in the case of Brownian
particles, admits as stationary state the distribution
$f=Z^{-1}\exp_{_{\{{\scriptstyle \kappa}\}}}[-(\beta mv^2/2-\mu)]$
which can be obtained also by maximizing the entropy
$S_{\kappa}=\int d^n v
\,[\,c(\kappa)f^{1+\kappa}+c(-\kappa)f^{1-\kappa}]$ with
$c(\kappa)=-Z^{\kappa}/\,[2\kappa(1+\kappa)]$ after properly
constrained.

\end {abstract}

\pacs{ PACS number(s): 05.10.Gg, 05.20.-y}

\section{Introduction}

In the last few decades there has been an intensive discussion on
non conventional classical or quantum statistics. For instance,
there are several experimental evidences of distributions with
tails which exhibit a power law decay. Up to now several entropies
with the ensuing statistics have been considered. A question which
arises spontaneously is if it is possible to obtain the stationary
statistical distribution of the various physical systems within
the framework of a time dependent scheme. The problem of the non
linear kinetics from a more general point of view in the
Fokker-Planck picture has been considered only in 1994. In ref.s
\cite{KQBF} an evolution equation has been proposed, describing a
generic non linear kinetics. Subsequently some properties of this
kinetics have been studied in ref.s \cite{NPB,PLA,RK}. \\
\\
$-------------------------$

\noindent ${}^*$  Email address: kaniadakis@polito.it
\\
%\noindent ${}^{}$ ${}^{}$ cond-mat/0103467

\noindent ${}^{}$ ${}^{}$ Accepted for publication on Phys. Lett.
A

\newpage
${}$ After 1995, in the Fokker-Planck picture, the anomalous
diffusion has been linked with the time dependent Tsallis
statistical distribution \cite{T,PP,TB,MPP,BO,BPPP,KL}. Finally,
the kinetics described by non linear Fokker-Planck equations has
been reconsidered in ref.s \cite{FD,F}.

Recently in the paper \cite{KU} it has been proposed the following
new, one-parameter deformation of the exponential and logarithm
functions:
\begin{eqnarray}
\exp_{_{\{{\scriptstyle \kappa}\}}}(x)&=&
\left(\sqrt{1+\kappa^2x^2}+\kappa x\right)^{1/\kappa} \ ,
\label{N00} \\ \ln_{_{\{{\scriptstyle \kappa}\}}}(x)&=&
\frac{x^{\kappa}-x^{-\kappa}}{2\kappa} \ . \label{N01}
\end{eqnarray}
These $\kappa$-deformed functions reduce to the standard
exponential and logarithm as the real deformation parameter
$\kappa$ approaches to zero and present a power law asymptotic
behaviour: $\exp_{_{\{{\scriptstyle
\kappa}\}}}(x){\atop\stackrel{\textstyle\sim}{\scriptstyle
x\rightarrow \pm \infty}}|2\kappa x|^{\pm 1/|\kappa|}$  and
$\ln_{_{\{{\scriptstyle
\kappa}\}}}(x){\atop\stackrel{\textstyle\sim}{\scriptstyle
x\rightarrow {\,0^+}}}-|2\,\kappa|^{-1}x^{-|\kappa|}$ $;\,\,\,$
$\ln_{_{\{{\scriptstyle
\kappa}\}}}(x){\atop\stackrel{\textstyle\sim}{\scriptstyle
x\rightarrow +\infty}}|2\,\kappa|^{-1}x^{|\kappa|}$.

We remark that the $\kappa$-exponential is the only function
besides the standard exponential, satisfying  the condition
$\exp_{_{\{{\scriptstyle \kappa}\}}}(-x)\exp_{_{\{{\scriptstyle
\kappa}\}}}(x)=1$ (both functions decrease for $x\rightarrow
-\infty$ and increase for $x\rightarrow +\infty$ with the same
rapidness) and obeying the scale law $[\exp_{_{\{{\scriptstyle
\kappa}\}}}(x)]^{\lambda}=\exp_{_{\{{\scriptstyle
\kappa'}\}}}(\lambda x)$. It is easy to verify that the standard
exponential can be obtained by posing $\kappa'=\kappa$ while the
$\kappa$-exponential given by Eq. (\ref{N00}) is found after
setting $\kappa'=\kappa/\lambda$.

In the same paper \cite{KU} a $\kappa$-kinetics has been
constructed which obeys the H-theorem and admits as stationary
state the distribution $f=Z^{-1}\exp_{_{\{{\scriptstyle
\kappa}\}}}[-(\beta mv^2/2-\mu)]$. This distribution can be
obtained also by maximizing, after properly constrained, the
entropy
\begin{equation}
S_{\kappa}=\int d^n v
\,[\,c(\kappa)f^{1+\kappa}+c(-\kappa)f^{1-\kappa}] \ ,
\end{equation}
being $c(\kappa)=-Z^{\kappa}/\,[2\kappa(1+\kappa)]$, which reduces
to the standard entropy as the deformation parameter approaches to
zero.

In the present paper, starting from the transition probability
which defines the most general non linear kinetics in the
Fokker-Planck picture, we obtain the associated generalized
entropy. Subsequently, after obtaining the statistical
distribution both as stationary solution of the non linear
evolution equation and as the state which maximizes the
generalized entropy, we demonstrate the H-theorem. Finally, we
reconsider some distributions already known in the literature in
the frame of the approach here proposed and in particular we
consider more extensively the $\kappa$-kinetics.

\section{Transition probabilities for the non linear kinetics}

 Let us consider a particle system interacting with its
environment which we consider as a bath. We denote with
$f=f(t,\mbox{\boldmath $v$})$ and $f'=f(t,\mbox{\boldmath $v$}')$
the particle densities in the sites $\mbox{\boldmath $v$}$ and
$\mbox{\boldmath $v$}'$ respectively, and postulate for the
transition probability $\pi(t,\mbox{\boldmath
$v$}\rightarrow\mbox{\boldmath $v$}')$
 from the site $\mbox{\boldmath $v$}$ to the
site $\mbox{\boldmath $v$}'$ the following form
\begin{equation}
\pi(t,\mbox{\boldmath $v$}\rightarrow\mbox{\boldmath $v$}')=
W(t,\mbox{\boldmath $v$},\mbox{\boldmath $v$}') \, \gamma(f,f') \
\ . \label{N1}
\end{equation}
The above transition probability is given by a product of two
factors. The first, $W(t,\mbox{\boldmath $v$},\mbox{\boldmath
$v$}')$ in (\ref{N1}) is the transition rate which depends on the
nature of the interaction between the particle and the bath and is
a function of the starting $\mbox{\boldmath $v$}$ and arrival
$\mbox{\boldmath $v$}'$ sites.

The second factor $\gamma(f,f')$ in (\ref{N1}) is an arbitrary
function of the particle populations of the starting and of
arrival sites. This function must satisfy the condition
$\gamma(0,f')=0 $ because, if the starting site is empty, the
transition probability is equal to zero. The dependence of the
function $\gamma(f,f')$ on the particle population $f'$ of the
arrival site plays a very important role in the particle kinetics
because can stimulate or inhibit the particle transition
$\mbox{\boldmath $v$}\rightarrow\mbox{\boldmath $v$}'$ in such a
way that interactions originated from collective effects can be
taken into account. The condition $\gamma(f,0) \neq 0$ requires
that in the case the arrival site is empty the transition
probability must depends only on the population of the starting
site. We note that for the standard linear kinetics the relation
$\gamma(f,f')= f$ holds.

In this paper, we will study kinetics coming from the transition
probabilities (\ref{N1}) when the function $\gamma$ satisfies the
condition
\begin{equation}
\frac{\gamma( f,f')}{\gamma( f',f)}=\frac{\kappa(f)}{\kappa(f')} \
\ , \label{N3}
\end{equation}
where $\kappa(f)$ is a positive real function. This condition
implies that $\gamma( f,f')/\kappa(f)$ is a symmetric function.
Then we can pose $\gamma( f,f')=\kappa(f)b(f)b(f')c(f,f')$ where
$b(f)$ and $c( f,f')=c(f',f)$ are two real arbitrary functions. It
will be convenient later on to introduce the real arbitrary
function $a(f)$ by means of
\begin{equation}
\kappa(f)=\frac{a(f)}{b(f)} \ \ , \label{N4}
\end{equation}
and write $\gamma( f,f')$ under the guise
\begin{equation}
\gamma(f,f')= a(f) \, b(f')\, c(f,f')\ \ . \label{N5}
\end{equation}
We claim at this point that $\gamma(f,f')$ given by (\ref{N5})
with $a(f)$ and $b(f')$ linked through (\ref{N4}), is the most
general function obeying the condition (\ref{N3}). We wish to note
that $\gamma(f,f')$ is given as a product of three factors. The
first factor  $a(f)$ is an arbitrary function of the particle
population of the starting site and satisfies the condition
$a(0)=0$ because if the starting site is empty the transition
probability is equal to zero. The second factor $b(f')$ is an
arbitrary function of the arrival site particle population. For
this function we have the condition $b(0)=1$ which requires that
the transition probability does not depend on the arrival site if,
in it, particles are absent. The expression of the function
$b(f')$ plays a very important role in the particle kinetics,
because it stimulates or inhibits the transition $\mbox{\boldmath
$v$}\rightarrow\mbox{\boldmath $v$}'$, allowing in such a way to
consider the interactions originated from collective effects.
Finally, the third factor $c(f,f')$ takes into account that the
populations of the two sites, namely $f$ and $f'$, can eventually
influence the transition, collectively and symmetrically. The
function $\gamma(f,f')$ given by (\ref{N5}) defines a special
interaction which involves, separately and/or together, the two
particle bunches entertained in the starting and arrival sites. We
recall that the interaction, depending on the relative distance of
the two involved sites, is taken into account in the transition
probability by the function $W$ defined in (\ref{N1}). The special
interaction defined through $\gamma(f,f')$ given by (\ref{N5}) can
be viewed as derived from a principle, the {\it Kinetical
Interaction Principle} (KIP). As we will see in the following
sections, the KIP governs the system evolving toward the
equilibrium and imposes the stationary state of the system.
Particular expressions of this principle are for instance the
Pauli exclusion principle \cite{KQF}, the generalized
exclusion-inclusion principle \cite{KQBF}, the Haldane generalized
exclusion principle \cite{NPB}, the principle underlying the
nonextensive statistics \cite{T,PP}.

\section{Fokker-Planck generalized kinetics}

We consider a classical stochastic marcoffian process in a
$n-$dimensional velocity space (of course the same process can be
considered in the physical space). It is described by the
distribution function $f=f(t,\mbox{\boldmath $v$})$ which obeys
the following evolution equation:

\begin{equation}
\frac{\partial f(t,\mbox{\boldmath $v$}) }{\partial t} =
\int_{\cal R} \left [ \pi(t,\mbox{\boldmath
$v$}'\rightarrow\mbox{\boldmath $v$})- \pi(t,\mbox{\boldmath
$v$}\rightarrow\mbox{\boldmath $v$}') \right ] \, d^n v' \ \ ,
\label{N6}
\end{equation}
where the transition probability according to the KIP is given by
$\pi(t,\mbox{\boldmath $v$}\rightarrow\mbox{\boldmath $v$}')=
W(t,\mbox{\boldmath $v$},\mbox{\boldmath $v$}') \, \gamma(f,f')$.
Let us write the transition rate as $W(t,\mbox{\boldmath
$v$},\mbox{\boldmath $v$}')=w(t,\mbox{\boldmath
$v$},\mbox{\boldmath $v$}'-\mbox{\boldmath $v$})$, where the
second argument in $w$ represents the change of the vector state
during the transition. For physical systems evolving very slowly,
$w(t,\mbox{\boldmath $v$},\mbox{\boldmath $v$}'-\mbox{\boldmath
$v$})$ decreases very expeditiously as $\mbox{\boldmath
$v$}'-\mbox{\boldmath $v$}$ increases. Then we can consider the
Kramers-Moyal expansion of Eq.(\ref{N6}):

\begin{eqnarray}
&&\frac{\partial f(t,\mbox{\boldmath $v$}) }{\partial t}=
\nonumber
\\&&\sum_{m=1}^{\infty}\Bigg[\frac{\partial^m
\{\zeta_{\alpha_1\alpha_2...\alpha_m}(t,\mbox{\boldmath $u$}) \,
\gamma[f(t,\mbox{\boldmath $u$}),f(t,\mbox{\boldmath
$v$})]\}}{\partial u_{\alpha_1}\partial u_{\alpha_2}...\partial
u_{\alpha_m}}+ \nonumber \\&&  (-1)^{m-1}
\zeta_{\alpha_1\alpha_2...\alpha_m}(t,\mbox{\boldmath
$v$})\frac{\partial^m \gamma[f(t,\mbox{\boldmath
$v$}),f(t,\mbox{\boldmath $u$})]}{\partial u_{\alpha_1}\partial
u_{\alpha_2}...\partial u_{\alpha_m}}\Bigg]_{_{\mbox{\boldmath
$u$}=\mbox{\boldmath $v$}}}  \label{N601}
\end{eqnarray}
where the $m$-th order momentum
$\zeta_{\alpha_1\alpha_2...\alpha_m}(t,\mbox{\boldmath $v$})$ of
the transition rate is defined as:
\begin{equation}
\zeta_{\alpha_1\alpha_2...\alpha_m}(t,\mbox{\boldmath
$v$})=\frac{1}{m!}\int_{\cal
R}y_{\alpha_1}y_{\alpha_2}...y_{\alpha_m} w(t,\mbox{\boldmath
$v$},\mbox{\boldmath $y$}) d^n y  .
\end{equation}
We remark that from Eq. (\ref{N601}) we can obtain as a particular
case Eq. (7) of ref. \cite{KQBF}. In the first neighbor
approximation only the first order (drift coefficient)
$\zeta_{i}(t,\mbox{\boldmath $v$})$ and the second order
(diffusion coefficient) $\zeta_{ij}(t,\mbox{\boldmath $v$})$
momenta of the transition rate are considered so that
Eq.(\ref{N601}) reduces to the following non linear second order
partial differential equation
\begin{eqnarray}
\frac{\partial f}{\partial t} = \frac{\partial}{\partial v_{i}}
\Bigg[&& \left (\zeta_{i}+\frac{\partial \zeta_{ij} } {\partial
v_{j}} \right )\gamma(f) \nonumber
\\ &&+\,\zeta_{ij}\,\gamma(f)\,\lambda(f)\, \frac{\partial f}{\partial v_{j}}
\,\Bigg ] \ \ , \label{N7}
\end{eqnarray}
with $\gamma(f)= \gamma( f, \,f)$ and
\begin{eqnarray}
\lambda(f)=\left[\frac{\partial}{\partial f}\ln \frac{\gamma( f,
f')} {\gamma(f',f)}\right]_{f'=f}
 \ \ . \nonumber
\end{eqnarray}
By taking into account the condition (\ref{N3}), the function
$\lambda(f)$ simplifies as
\begin{equation}
\lambda(f)=\frac{\partial \ln \kappa(f)}{\partial f}  \ \ ,
\end{equation}
and Eq.(\ref{N7}) becomes
\begin{eqnarray}
 \frac{\partial f}{\partial t} = \frac{\partial}{\partial
v_{i}}  \Bigg[\gamma(f)
 \left (\zeta_{i}+\frac{\partial \zeta_{ij} }
{\partial v_{j}} \right )\nonumber \\ +\gamma(f) \frac{\partial
\ln \kappa(f)}{\partial f}\,\zeta_{ij} \frac{\partial f}{\partial
v_{j}} \! \Bigg ] . \label{N8}
\end{eqnarray}

We assume the independence of motion among the $n$ directions of
the homogeneous and isotropic $n$-dimensional velocity space and
pose $\zeta_{i}=J_i$, $\zeta_{ij}=D\delta_{ij}$, being
$\mbox{\boldmath $J$}=\mbox{\boldmath $J$}(\mbox{\boldmath $v$})$
and $D=D(\mbox{\boldmath $v$})$ the drift and diffusion
coefficients respectively. Moreover we introduce the potential
$U=U(\mbox{\boldmath $v$})$ by means of
\begin{equation}
\beta \frac{\partial U}{\partial \mbox{\boldmath$v$}} =
\frac{1}{D} \left (\mbox{\boldmath$J$}+ \frac{\partial D }
{\partial \mbox{\boldmath$v$}} \right ) \ \ ,
\end{equation}
with $\beta$ a constant. Then we can write Eq. (\ref{N8}) in the
form
\begin{equation}
\frac{\partial f}{\partial t} = \frac{\partial}{\partial
\mbox{\boldmath$v$}} \! \left\{ D\gamma(f)\frac{\partial}{\partial
\mbox{\boldmath$v$}} \left [ \beta(U-\mu')+\ln \kappa(f) \right ]
\! \right \} , \label{N9}
\end{equation}
where $\mu'$ is an arbitrary constant. Eq. (\ref{N9}) can be
written also as
\begin{eqnarray}
\frac{\partial f}{\partial t} \,&& = \frac{\partial}{\partial
\mbox{\boldmath$v$}} \Bigg \{ D \,a(f)\,b(f)\,c(f) \nonumber \\
&&\times \frac{\partial}{\partial \mbox{\boldmath$v$}} \left [
\beta(U\!\!-\!\!\mu')+\ln \frac{a(f)}{b(f)} \right ] \! \Bigg \},
\label{N901}
\end{eqnarray}
with $c(f)=c(f,f)=\lim_{v'\rightarrow v} c[f(
\mbox{\boldmath$v$}),f( \mbox{\boldmath$v$}')]$. From this
definition we have that $c(f)$ can depend also on the derivative
of $f$. For instance we can choose $c(f)=C[f,(\partial
f/\partial\mbox{\boldmath$v$})^2]$ and so on. We note that Eq.
(\ref{N901}) in the case $c(f)=1$ reduces to Eq. (10) of ref.
\cite{NPB}.

In the following part of this section we will refer to the form
(\ref{N9}) for the evolution equation of the distribution function
$f(t,\mbox{\boldmath $v$})$. First of all we observe that this
equation is a non linear continuity equation, namely
\begin{equation}
\frac{\partial f}{\partial t} + \frac{\partial
\mbox{\boldmath$j$}}{\partial \mbox{\boldmath$v$}}=0 \ \ ,
\end{equation}
where the current
$\mbox{\boldmath$j$}=\mbox{\boldmath$j$}(\mbox{\boldmath$v$},f)$
assumes the form
\begin{equation}
\mbox{\boldmath$j$}=D \gamma(f)\,\mbox{\boldmath${\cal U}
$}(\mbox{\boldmath $v$},f) \ \ ,
\end{equation}
and the field  $\mbox{\boldmath${\cal U} $}(\mbox{\boldmath
$v$},f)$ can be derived from the potential $\Phi(\mbox{\boldmath
$v$},f)=-\beta(U-\mu') - \ln \kappa (f)$:
\begin{equation}
\mbox{\boldmath${\cal U}$}(\mbox{\boldmath $v$},f)= \frac{\partial
\Phi(\mbox{\boldmath $v$},f)}{\partial \mbox{\boldmath$v$}}  \ \ .
\end{equation}
It is easy to verify that this potential can be written as the
functional derivative $\Phi(\mbox{\boldmath $v$},f)=\delta {\cal
K}/\delta f$, of the functional ${\cal K}$ defined as
\begin{equation}
{\cal K}=- \int_{\cal R} d^n v \left[\int \ln \kappa (f) \  d f +
  \beta (U-\mu')f\right]  \ \ .
\end{equation}
The current $\mbox{\boldmath$j$}$ becomes
\begin{equation}
\mbox{\boldmath$j$} = D\gamma(f)\frac{\partial} {\partial
\mbox{\boldmath$v$}} \frac{\delta {\cal K}}{\delta f} \ \ ,
\end{equation}
and the evolution equation (\ref{N9}) of the distribution function
$f(t,\mbox{\boldmath$v$})$ assumes the following compact form:
\begin{equation}
\frac{\partial f}{\partial t} + \frac{\partial}{\partial
\mbox{\boldmath$v$}} \! \left[D\gamma(f)\frac{\partial} {\partial
\mbox{\boldmath$v$}} \frac{\delta {\cal K}}{\delta f} \right ]=0 \
\ . \label{N10}
\end{equation}
We observe that from the structure of the evolution equation
(\ref{N10}) the stationary distribution $f_{\!s}=f(\infty,
\mbox{\boldmath$v$})$ of the system can be obtained from a
variational principle:
\begin{eqnarray}
\frac{\delta {\cal K}}{\delta f}=0 \ \ \ \Rightarrow \ \ \
\frac{\partial f}{\partial t}=0 \ \ ; \ \ f=f_{\!s} \ \ ,
\nonumber
\end{eqnarray}
and assumes the form:
\begin{equation}
\ln \kappa (f_{\!s})=-\beta(U-\mu') \ \ , \label{SS}
\end{equation}
where the constant $\mu'$ will be calculated taking into account
the normalization condition $\int_{\cal R}d^n v \, f=1$. We
consider the time evolution of the functional ${\cal K}={\cal
K}(t)$ which can write in the form:
\begin{equation}
{\cal K}= \int_{\cal R}\ d^n v \, K(f,f_{\!s}) \ \ , \label{M1}
\end{equation}
where its density is given by
\begin{equation}
 K(f,f_{\!s})= - \int df  \ln \frac{\kappa (f)}{\kappa
(f_{\!s})}  \ \ . \label{M2}
\end{equation}

\section{H-theorem within the framework of the non linear kinetics}

The time derivative of ${\cal K}$ is given by
\begin{equation}
\frac{d{\cal K}}{d t}= \int_{\cal R} d^n v \frac{\delta {\cal
K}}{\delta f} \frac{\partial f}{\partial t}
 \ \ ,
\end{equation}
where $\partial f /\partial t$ can be calculated from (\ref{N10}):
\begin{eqnarray}
\frac{d{\cal K}}{d t}= -\int_{\cal R} d^n v \frac{\delta {\cal
K}}{\delta f}\frac{\partial}{\partial \mbox{\boldmath$v$}} \!
\left[D\gamma(f)\frac{\partial} {\partial \mbox{\boldmath$v$}}
\frac{\delta {\cal K}}{\delta f} \right ] \ \ . \nonumber
\end{eqnarray}
Performing an integration by parts and assuming the appropriate
boundary conditions for the current, we obtain
\begin{equation}
\frac{d{\cal K}}{d t}= \int_{\cal R} d^n v
D\gamma(f)\left(\frac{\partial}{\partial \mbox{\boldmath $v$}}
\frac{\delta {\cal K}}{\delta f} \right)^2 \ \ .
\end{equation}
We may conclude that ${\cal K}$ is an increasing function
\begin{equation}
\frac{d{\cal K}}{d t}\geq 0 \ \ .\label{N11}
\end{equation}
In order to study the behaviour of ${\cal K}$ when
$t\rightarrow\infty$ we introduce the function
\begin{equation}
\sigma(f)= - \int d f \, \ln \kappa (f)   \ \ ,\label{M6}
\end{equation}
so that $\kappa (f)$ can be written as
\begin{equation}
\kappa (f)=\exp\left[-\frac{d \sigma(f)}{d f}\right] \ \ .
\end{equation}
Now we are able to calculate, in the limit $t\rightarrow\infty$,
the following difference
\begin{eqnarray}
&&{\cal K}(t)\!-\!{\cal K}(\infty) \nonumber \\&& =\int_{\cal R}
\!d^nv \ [\sigma(f)-\sigma(f_{\!s})+(f-f_{\!s})\ln \kappa
(f_{\!s})] \nonumber \\ && = \int_{\cal R}
d^nv\left[\sigma(f)-\sigma(f_{\!s})-(f-f_{\!s}) \frac{d \sigma
(f_{\!s})}{d f_{\!s} }\right] \nonumber \\ &&  \approx  \int_{\cal
R} d^nv\left[\frac{1}{2}\frac{d^2 \sigma (f_{\!s})}{d f^2_{\!s}
}(f-f_{\!s})^2 \right] \ ,
\end{eqnarray}
and assume that $d^2 \sigma (f)/d f^2  \leq 0$. This requirement
is satisfied if  the function $\kappa(f)$ obeys to the condition
\begin{equation}
\frac{d \kappa (f)}{d f } \geq 0\ \ .
\end{equation}
Consequently we have:
\begin{eqnarray}
{\cal K}(t)\leq {\cal K}(\infty) \ \ . \label{N12}
\end{eqnarray}

We discuss now briefly the meaning of ${\cal K}$. Firstly, we
observe that Eq.s (\ref{N11}) and (\ref{N12}) constitute the
H-theorem for the non linear system governed by the evolution
equation (\ref{N10}) and implies that $-{\cal K}$ is a Lyapunov
functional. Secondly, the functional ${\cal K}$ given by
(\ref{M1}) and (\ref{M2}) can be written as the sum of two terms
${\cal K}=S + S_c$. The first term is the free system entropy
\begin{equation}
 S=  \int_{\cal R} d^n v \, \, \sigma(f) \ \ ,
\end{equation}
while the second term, $S_c=-\beta (E-\mu')$, represents an
entropy originated from the constraints imposed by the
normalization requirement, and by the mean value of the relevant
energy of the system, defined as $E= \int_ {\cal R} d^n v \ U f$.
With these positions ${\cal K}$ can be viewed as the entropy of
the constrained system, or alternatively, as the entropy of the
universe: system + environment. We observe on the other hand that,
by deriving with respect to $t$ the equation ${\cal K}=S + S_c$
and taking into account the expression of $S_c$, one obtains
immediately the Clausius inequality:
\begin{equation}
\frac{d S}{d t}\geq \beta\, \frac{d E}{d t} \ \ .
\end{equation}
Finally, the definition (\ref{M1})-(\ref{M2}) and the
relationships (\ref{N11}) and (\ref{N12}) suggest that the
quantity
\begin{equation}
{\cal D}(t)={\cal K}(\infty)-{\cal K}(t) \ \ ,
\end{equation}
can be viewed as the statistical distance between the two
distributions $f_{\!s}$ and $f$. The maximization of the
constrained entropy ${\cal K}$ implies the minimization of the
distance ${\cal D}(t)$. This distance ${\cal D}(t)$, in the case
of the linear kinetics where $\kappa(f)=f$, reduces to the well
known Kullback-Leibler distance \cite{PMP,RCP}:
\begin{equation}
 {\cal D}(t)=  \int_{\cal R} d^n v \, \, f \ln
 \frac{f}{f_{\!s}} \ \ .
\end{equation}

In concluding this section we remark that, starting from the
function $K=K(f,f_{\!s})$ given by (\ref{M2}) we can consider the
class of functionals defined through:
\begin{equation}
\Lambda = \chi_1\left(  \int_ {\cal R} d^n
v\,\,\chi_2\left(K\right)\right) \ \ , \label{N13}
\end{equation}
being $\chi_1$ and $\chi_2$ two arbitrary algebraic functions.
After indicating with ${\chi_1}'$ and ${\chi_2}'$ their first
derivatives, one  obtains the following expression for the
functional derivative of $\Lambda$
\begin{equation}
\frac{\delta \Lambda}{\delta f}={\chi_1}'
 \left( \int_ {\cal R}
d^n v\,\,\chi_2(K)\right) {\chi_2}'(K)\frac{\delta {\cal
K}}{\delta f} \ \ .
\end{equation}
It is apparent that if we choose $\chi_1$ and $\chi_2$ with the
appropriate properties, $-\Lambda$ can be a Lyapunov functional.
The quantity $\Lambda$ can be chosen as the entropic functional
and then can replace ${\cal K}$. We recall that, for instance, the
Renyi \cite{R} and the Sharma-Mittal \cite{SM} entropies, can be
obtained as particular cases of (\ref{N13}).

\section{Quantum statistics}

In this section let us report Fokker-Planck equations for few
quantum distributions to illustrate the relevance of the approach
previously proposed which permits to treat the various non linear
kinetics in a unifying scheme. In the following we consider the
case of brownian particles where $U=mv^2/2$ and $D=const$.

As first example we consider the kinetics defined through
$a(f)=f$, $b(f)=1 +\eta f$ and $c(f)=1$. We obtain
$\kappa(f)=f/(1+\eta f)$ and $\gamma(f)=f(1+\eta f)$ while the
evolution equation Eq. (\ref{N9}) becomes:
\begin{equation}
\frac{\partial f}{\partial t} = \frac{\partial}{\partial
\mbox{\boldmath$v$}} \! \left[D\beta m\mbox{\boldmath$v$}f(1+\eta
f)+D\frac{\partial f}{\partial \mbox{\boldmath$v$}} \! \right ] \
\ , \label{N24}
\end{equation}
which it has been previously proposed in  ref. \cite{KQBF}.

The stationary state of this equation is the quantum statistical
distribution $f=(\exp \epsilon -\eta)^{-1}$ being
$\epsilon=\beta\,(mv^2/2-\mu)$. For $\eta=0$ we obtain the
classical Maxwell-Boltzmann statistics and Eq. (\ref{N24}) reduces
to the linear Fokker-Planck equation. For which for $\eta=-1$, the
factor $b(f)$ takes into account the Pauli exclusion principle and
the stationary distribution define the Fermi-Dirac statistics. For
$\eta=1$ an inclusion principle holds and  we obtain the
Bose-Einstein statistics. Finally for $\eta\not=\pm1$ we have a
intermediate quantum statistics interpolating between the Bose and
Fermi ones.

The boson-like $(\eta=1)$ or fermion-like $(\eta=-1)$ quon
statistics \cite{PLA} can be obtained easily by posing
$a(f)=[f]_q$ and $b(f')=[1+\eta f']_q$, where
$[x]_{_{\,\scriptstyle q}}=(q^{x}-q^{-x})/2\ln q$ and $q\in {\bf
R}$. The stationary distribution is given through $[f]_q/[1+\eta
f]_q=\exp (-\epsilon)$. If we choose for simplicity
$c(f)=c_q=(2\ln q) / (q-q^{-1})$, the evolution equation becomes
\cite{PLA}:
\begin{equation}
\frac{\partial f}{\partial t} = \frac{\partial}{\partial
\mbox{\boldmath$v$}} \! \left(c_{q}\,D\beta
m\mbox{\boldmath$v$}\,[f]_{_{\,\scriptstyle q}}[1+\eta
f]_{_{\,\scriptstyle q}}+D \frac{\partial f}{\partial
\mbox{\boldmath$v$}} \, \right ) \ \ .
\end{equation}

\section{Nonextensive statistics}

In this section we consider two examples of classical non linear
kinetics. After considering briefly the non linear kinetics
admitting as stationary distribution the Maxwell-Boltzmann one, we
introduce the nonextensive Tsallis kinetics within the framework
of the theory developed in the sections II, III and IV.

We write Eq.(\ref{SS}) in the case of brownian particles as
follows

\begin{equation}
\kappa(f_s)=\frac{1}{Z}\exp\left[ -\beta\left(
\frac{1}{2}mv^2-\mu\right)\right] \label{FSS} \ \ ,
\end{equation}
where $Z$ is calculated from the condition $ \int_{\cal R} d^n v
f_s=1 $. We remark that $f_s$ given by Eq. (\ref{FSS}), can be
obtained from the variational principle
\begin{eqnarray}
\frac{\delta}{\delta f} \int_{\cal R} d^n v \Bigg [-\int \ln
\kappa (f) d f - \beta \frac{1}{2}mv^2f && \nonumber \\
   + (\beta\mu-\ln
Z) f && \Bigg]=0   \ \label{VAR} ,
\end{eqnarray}
and also as stationary solution of the evolution equation:
\begin{equation}
 \frac{\partial f}{\partial t} = \frac{\partial}{\partial
\mbox{\boldmath$v$}} \! \left\{D\, \gamma(f)\frac{\partial
}{\partial \mbox{\boldmath$v$}}\Bigg[\ln \kappa(f)-\ln\kappa(f_s)
\Bigg] \right \} .
\end{equation}

{\it Maxwell-Boltzmann statistics:} We start by considering the MB
statistics given by $f_s= Z^{-1} \exp (- \beta m v^2/2)$. It is
readily seen that the related kinetics is defined posing $a(f)=f$,
$b(f)=1$, while the symmetric function $c(f)$ remains arbitrary.
Then we have $\kappa(f)=f$ and $\gamma(f)=f c(f)$. The evolution
equation becomes
\begin{equation}
\frac{\partial f}{\partial t} = \frac{\partial}{\partial
\mbox{\boldmath$v$}} \! \left\{D\, \gamma(f)\frac{\partial
}{\partial \mbox{\boldmath$v$}}\Bigg[\ln \Big(f/f_s\Big) \Bigg]
\right \} \ \ . \label{LE}
\end{equation}
We observe that there are infinite ways, one for any choice of
$c(f)$, to obtain the MB distribution. This statistical
distribution can be obtained also after maximizing the standard
additive (extensive) Boltzmann-Gibbs-Shannon entropy
\begin{equation}
 S=  -\int_{\cal R} d^n v \, \, f \ln f \ \ ,
\end{equation}
after properly constrained (we set $k_B=1$).

{\it Tsallis statistics:} We consider the non extensive
termostatistics introduced by Tsallis \cite{T} which can be
obtained naturally in the present frame after choosing properly
$\kappa (f)$ while $\gamma(f)$ remains an arbitrary function.

\noindent A) Let us consider the  kinetics defined by fixing
$\kappa(f)$ through
\begin{equation}
\ln[Z\kappa(f)]= \ln_q(Zf) \ \ ,
\end{equation}
where $\ln_q(x)=(x^{1-q}-1)/(1-q)$. We note that $d\kappa(f)/df
\geq 0$ for $\forall q \in R$.  The kinetics of the system is
described by means of the evolution equation:
\begin{equation}
\frac{\partial f}{\partial t} = \frac{\partial}{\partial
\mbox{\boldmath$v$}} \! \left\{D\, \gamma(f)\frac{\partial
}{\partial \mbox{\boldmath$v$}}\Big[\ln_q(Zf)- \ln_q(Zf_s) \Big]
\! \right \} \ \ , \label{N18}
\end{equation}
which, in the case $\gamma(f)=f$, reproduces the equation proposed
in ref. \cite{PP}. The stationary solution of Eq. (\ref{N18}) is
given by:
\begin{equation}
f = \frac{1}{Z}\exp_q \left[-\beta\left(\frac{1}{2} m
v^2-\mu\right)\right] \ \ , \label{N191}
\end{equation}
with $\exp_q(x)=[1+(1-q)x]^{1/(1-q)}$ and $Z=\int_{\cal R} d^n v
\,\exp_q \left[-\beta\left( m v^2/2-\mu\right)\right]$. This
distribution can be obtained also from the variational principle
given by Eq. (\ref{VAR}):
\begin{eqnarray}
\frac{\delta}{\delta f} \int_{\cal R} d^n v \Bigg[-\frac{1}{2-q}
\frac{(Zf)^{1-q}-1}{1-q}+\frac{1}{2-q} \nonumber  \\ - \beta
\frac{1}{2}mv^2+ \beta\mu \Bigg]f=0 \ \label{VART1}  \ .
\end{eqnarray}
From Eq. (\ref{VART1}) it results apparent that the entropy is
given by:
\begin{equation}
S_q [f] = -\frac{1}{2-q}<\ln_q(Zf)>+\frac{1}{2-q} \ \label{ENTRT1}
\ .
\end{equation}
We recall that in the frame of the present non linear kinetics the
mean value of $A$ is defined through
\begin{equation}
<A>=\int_{\cal R} d^n v A f \ \ \ ; \ \ \ \int_{\cal R} d^n v f=1.
\label{Norma}
\end{equation}

\noindent B) Taking into account that
$\exp_q(-x)\neq[\exp_q(x)]^{-1}$, we consider the statistical
distribution
\begin{equation}
f = \frac{1}{Z}\left\{\exp_q \left[\beta\left(\frac{1}{2} m
v^2-\mu\right)\right]\right\}^{-1} \ \ , \label{N192}
\end{equation}
which can be viewed as the steady state of the kinetics defined by
fixing $\kappa(f)$ through $\ln[Z\kappa(f)]= -\ln_q(1/Zf)$. Also
in this case we have $d\kappa(f)/df \geq 0$ for $\forall q \in R$.
The relation $[\exp_q(x)]^{-1}=\exp_{2-q}(-x)$ imposes that
(\ref{N192}) can be obtained by maximizing the entropy
\begin{equation}
S_{2-q}[f] = -\frac{1}{q}<\ln_{2-q}(Zf)>+\frac{1}{q} \
\label{ENTRT2} \ \ ,
\end{equation}
after properly constrained.

\noindent C) The relation $[\exp_{q}(x)]^{\,q}=\exp_{2-1/q}(q x)$
suggests to write the statistical distribution under the form:
\begin{equation}
f=p^q \ \ \ \ ; \ \ \ \ p = \frac{1}{Z}\exp_q
\left[-\beta\left(\frac{1}{2} m v^2-\mu\right)\right] \ .
\label{N193}
\end{equation}
This distribution can be viewed as the steady state of the
kinetics defined by imposing $\ln[Z\kappa(f)]= \ln_q(Zf^{1/q})$.
We remark that $d\kappa(f)/df \geq 0$ for $q > 0$. In the
following we consider $q > 0$ in order to keep the H-theorem. Of
course the normalization condition $\int_{\cal R} d^n v \,f=1$
implies that:
\begin{equation}
\int_{\cal R} d^n v \,p^q =1 \ \ \ \ ; \ \ \ \ \int_{\cal R} d^n v
\,p \neq 1 \ \ , \label{NORM}
\end{equation}
and
\begin{equation}
Z^q = \int_{\cal R} d^n v \,\left\{\exp_q \left[-\beta\left( m
v^2/\,2-\mu\right)\right]\right\}^q \ \ . \label{ZETA}
\end{equation}
The mean value of $A$ can be calculated as:
\begin{eqnarray}
<A>&&=\frac{\int_{\cal R} d^n v \,A\, f}{\int_{\cal R} d^n v \,f}=
\frac{\int_{\cal R} d^n v \,A\, p^q}{\int_{\cal R} d^n v \, p^q}
\nonumber \\ &&= \int_{\cal R} d^n v \,A\, p^q \ \ .
\end{eqnarray}
The variational principle expressed by Eq. (\ref{VAR}) in the case
of the distribution function $p=f^{1/q}$ can be written as
\begin{eqnarray}
\frac{\delta}{\delta p}\,\, \int_{\cal R} d^n v  \Big[&&\,
\frac{1-q\,(Z p )^{1-q}}{1-q} \nonumber \\ && \,- \beta
\frac{1}{2}mv^2+ \beta\mu\Big]p^{\,q}=0 \ \label{VART3}  \ ,
\end{eqnarray}
and the entropy consequently is defined through:
\begin{equation}
S_q[p]= -q<\ln_q(Zp)> +1 \label{ENTRT3} \ .
\end{equation}

We conclude by remarking that here we have obtained Tsallis
statistics, within the framework of a general non linear kinetics,
in a new version, different from the ones already known in
literature. We note that the present version in terms of escort
probabilities (\ref{N193})-(\ref{ENTRT3}) is consistent with the
proposal of ref. \cite{QW} where the validity of Eq.s (\ref{NORM})
has been postulated to solve some open problems in Tsallis
statistics.

\section{The \boldmath$\kappa$-statistics}

Let us consider a deformation of the logarithm, namely
$\ln_{_{\{{\scriptstyle \kappa}\}}}(f)$,  obeying the condition
$\ln_{_{\{{\scriptstyle \kappa}\}}}(f^{-1})=
-\ln_{_{\{{\scriptstyle \kappa}\}}}(f)$. The most general solution
of this last equation is given by $\ln_{_{\{{\scriptstyle
\kappa}\}}}(f)=[\lambda_{\kappa}(f)-\lambda_{\kappa}(f^{-1})]/2$
being $\lambda_{\kappa}(f)$ a real arbitrary function. Let us
consider only one parameter $(\kappa)$ deformations and impose
that $\ln_{_{\{{\scriptstyle \kappa}\}}}(f)$ obeys to the
condition $\ln_{_{\{{\scriptstyle
\kappa}\}}}(f^m)=m\ln_{_{\{{\scriptstyle \kappa'}\}}}(f)$, being
$\kappa'=\kappa'(m,\kappa)$. It is easy to verify that two
solutions exist. First trivial solution is: $\kappa'=\kappa$,
$\ln_{_{\{{\scriptstyle \kappa}\}}}(f)=\ln (f)$. Second solution
is given by: $\kappa'=m\kappa$, $\ln_{_{\{{\scriptstyle
\kappa}\}}}(f)=(f^{\kappa}-f^{-\kappa})/2\kappa$. We note that
this one parameter family of solutions contains, as limiting case
for $\kappa \rightarrow 0$, the trivial undeformed logarithm. The
inverse function of the $\kappa$-logarithm is the
$\kappa$-exponential, namely $f=\exp_{_{\{{\scriptstyle
\kappa}\}}}(x)$. Very recently  in ref. \cite{KU}, starting from
this deformed exponential function a one-parameter deformed
mathematics has been constructed which shows various very
interesting symmetries.

In the same reference the $\kappa-$kinetics has been proposed and
studied. This deformed kinetics can be introduced naturally by
posing that $\gamma(f)$ remains a arbitrary function, while
$\kappa (f)$ is given by
\begin{equation}
\ln[Z\kappa(f)]=\ln_{_{\{{\scriptstyle \kappa}\}}}(Zf) \ \ .
\end{equation}
We remark that $d\kappa(f)/df \geq 0$ for $\forall \kappa \in R$
so that the H-theorem still holds. The evolution equation becomes
\begin{equation}
\frac{\partial f}{\partial t} = \frac{\partial}{\partial
\mbox{\boldmath$v$}} \! \left\{D\, \gamma(f)\frac{\partial
}{\partial \mbox{\boldmath$v$}}\Big[ \ln_{_{\{{\scriptstyle
\kappa}\}}}(Zf)-\ln_{_{\{{\scriptstyle \kappa}\}}}(Zf_s)\Big] \!
\right \} \ \ , \label{N20}
\end{equation}
and its stationary distribution is given by
\begin{equation}
f_s = \frac{1}{Z}\exp_{_{\{{\scriptstyle \kappa}\}}}
\!\left[-\beta\left(\frac{1}{2} m v^2-\mu\right)\right] \ \ ,
\label{N21}
\end{equation}
which, after calculation of $Z$, in the case $\mu=0$, assumes the
form
\begin{eqnarray}
f_s\,\, &&=
\left[\frac{\beta\,m\,|\kappa|}{\pi}\right]^\frac{n}{2}\left
[1+\frac{n}{2}\,|\kappa|\right] \nonumber \\ && \times \,
\frac{\Gamma\left(\frac{1}{2\, |\kappa|}+
\frac{n}{4}\right)}{\Gamma\left(\frac{1}{2\,|\kappa|}-\frac{n}{4}
\right)}\, \exp_{_{\{{\scriptstyle \kappa}\}}}\!\left(
-\frac{\beta}{2}\,m\,v^2\right) \ ,\label{N03}
\end{eqnarray}
with \hbox{$|\kappa|<2/n$} and \hbox{$n=1,2,3$} being  the
dimension of the velocity space. The asymptotic behaviour of the
distribution $f_s$ of (\ref{N03}) is
\begin{eqnarray}
f_s \, {\atop\stackrel{\textstyle\sim}{\scriptstyle v\rightarrow
+\infty}}\,&&\pi^{-\frac{n}{2}}
\Big(\beta\,m\,|\kappa|\Big)^{\frac{n}{2}-
\frac{1}{|\kappa|}}\left [1+\frac{n}{2}\,|\kappa|\right] \nonumber
\\ && \times \, \frac{\Gamma\left(\frac{1}{2\, |\kappa|}+
\frac{n}{4}\right)}{\Gamma\left(\frac{1}{2\,|\kappa|}-\frac{n}{4}
\right)}\,\, v^{-\frac{2}{|\kappa|}} \ \ . \label{N211}
\end{eqnarray}
The $r$-order momentum
\begin{equation}
<\!\!v^r\!\!>_{\kappa} \, = \! \frac{\int_{\cal R} d^nv\, \, v^r
\, f}{\int_{\cal R} d^nv\,\, f} \ \ ,
\end{equation}
of the distribution function (\ref{N03}) is finite when
$|\kappa|<2/(n+r)$ and is given by

\begin{eqnarray}
<\!\!v^r\!\!>_{\kappa} \, \, = \left(\,m\,\beta
\,|\kappa|\,\right)^{-\frac{r}{2}}
\frac{1+\frac{n}{2}\,|\kappa|}{1+\frac{n+r}{2}\,|\kappa|}\,
\,\,\frac{\Gamma\left(\frac{n+r}{2}\right)}{\Gamma\left(\frac{n}{2}
\right)} && \, \nonumber \\ \times \,
\frac{\Gamma\left(\frac{1}{2\,|\kappa|}+
\frac{n}{4}\right)}{\Gamma\left(\frac{1}{2\,|\kappa|}+\frac{n+r}{4}
\right)}\,\,\frac{\Gamma\left(\frac{1}{2\,|\kappa|}-
\frac{n+r}{4}\right)}{\Gamma\left(\frac{1}{2\,|\kappa|}-\frac{n}{4}
\right)}&& \ .\label{N311}
\end{eqnarray}

It is easy to verify that the distribution (\ref{N03}) can be
obtained after maximization under the appropriate constrains of
the entropy
\begin{equation}
S_{\kappa}=-\frac{1}{2\kappa}\,\int_{\cal R} d^nv \,\left(\,
\frac{Z^{\,\kappa}}{1\!+\!\kappa}\, f^{\,1+\kappa}-
\frac{Z^{\!-\kappa}}{1\!-\!\kappa}\, f^{\,1-\kappa} \right)\ ,
\label{N300}
\end{equation}
which reduced to the standard Shannon entropy $S_0=-\int d^nv \, f
\ln Zf $ as the the deformation parameter $\kappa\rightarrow 0$.
We note that the entropy $S_{\kappa}$ and the entropy $S=-
\int_{\cal R} d^n v \, \, \int d f \, \ln \kappa (f)$ are
connected through $S_{\kappa}=S-\ln Z$, being $\int d^nv \, f=1$.
The variational equation (\ref{VAR}) becomes:

\begin{equation}
\frac{\delta}{\delta f}\left[S_{\kappa} + \int_{\cal R} d^n v
\left(
 - \beta \frac{1}{2}mv^2+ \beta\mu\right)f\right]=0
\ \label{VART} \ .
\end{equation}

Let us define, for real positive functions, the following
$\kappa$-product:
\begin{eqnarray}
f\otimes\mbox{\raisebox{-2.5mm}{\hspace{-3.5mm}$\scriptstyle
\kappa$}} \hspace{2mm}g= \exp_{_{\{{\scriptstyle
\kappa}\}}}\!\left(\,\ln_{_{\{{\scriptstyle
\kappa}\}}}(f)+\ln_{_{\{{\scriptstyle \kappa}\}}}(g)\right) \ ,
\label{N04}
\end{eqnarray}
which reduces to the ordinary product as $\kappa\rightarrow 0$,
namely $f \otimes\mbox{\raisebox{-2.5mm}
{\hspace{-3.5mm}$\scriptstyle \, 0$}} \hspace{2mm}g= f\cdot g$.
The above $\kappa$-product has the same properties of the ordinary
one: i) associative law:
$(f\otimes\mbox{\raisebox{-2.5mm}{\hspace{-3.5mm}$\scriptstyle\kappa$}}
\hspace{2mm}g)\otimes\mbox{\raisebox{-2.5mm}{\hspace{-3.5mm}$
\scriptstyle\kappa$}}\hspace{2mm}h=f\otimes\mbox{\raisebox{-2.5mm}
{\hspace{-3.5mm}$\scriptstyle\kappa$}}\hspace{2mm}
(g\otimes\mbox{\raisebox{-2.5mm}{\hspace{-3.5mm}$\scriptstyle\kappa$}}
\hspace{2mm}h)$; ii) neutral element:
$f\otimes\mbox{\raisebox{-2.5mm}{\hspace{-3.5mm}$\scriptstyle\kappa$}}
\hspace{2mm}1=1\otimes\mbox{\raisebox{-2.5mm}{\hspace{-3.5mm}$
\scriptstyle\kappa$}}\hspace{2mm}f= f$; iii) inverse element:
$f\otimes\mbox {\raisebox{-2.5mm}
{\hspace{-3.5mm}$\scriptstyle\kappa$}}\hspace{2mm}(1/f)=
(1/f)\otimes\mbox {\raisebox{-2.5mm}
{\hspace{-3.5mm}$\scriptstyle\kappa$}}\hspace{2mm}f=1$; iv)
commutative law : $f \otimes\mbox{\raisebox{-2.5mm}
{\hspace{-3.5mm}$\scriptstyle\kappa$}}
\hspace{2mm}g=g\otimes\mbox{\raisebox{-2.5mm}
{\hspace{-3.5mm}$\scriptstyle\kappa$}}\hspace{2mm}f$. Of course,
the $\kappa$-division can be defined as
$f\oslash\mbox{\raisebox{-2.5mm}{\hspace{-3.5mm}$\scriptstyle
\kappa$}} \hspace{2mm}g=f
\otimes\mbox{\raisebox{-2.5mm}{\hspace{-3.5mm} $\scriptstyle
\kappa$}}\hspace{2mm} (1/g)$. Finally,
$f\otimes\mbox{\raisebox{-2.5mm}{\hspace{-3.5mm}$\scriptstyle\kappa$}}
\hspace{2mm}0=0\otimes\mbox{\raisebox{-2.5mm}{\hspace{-3.5mm}$
\scriptstyle\kappa$}}\hspace{2mm}f= 0$.  We remark the following
interesting property of the $\kappa$-exponential:
\begin{eqnarray} \exp_{_{\{{\scriptstyle
\kappa}\}}}(x)\otimes\mbox{\raisebox{-2.5mm}{\hspace{-3.5mm}
$\scriptstyle \!\! \kappa$}}\hspace{2mm} \!\exp_{_{\{{\scriptstyle
\kappa}\}}}(y)=\exp_{_{\{{\scriptstyle \kappa}\}}}(x+y) \ .
\label{N05}
\end{eqnarray}
Eq. (\ref{N05}) in the case $y=-x$ reduces to the relationship
$\exp_{_{\{{\scriptstyle \kappa}\}}}(x) \exp_{_{\{{\scriptstyle
\kappa}\}}}(-x)=1$, which is the starting point of ref. \cite{KU}
for the construction of the $\kappa$-exponential.

We conclude the discussion by noting that the evolution equation
(\ref{N20}) can be written in the form

\begin{equation}
\frac{\partial f}{\partial t} = \frac{\partial}{\partial
\mbox{\boldmath$v$}} \! \left(D\, \gamma(f)\frac{\partial
}{\partial \mbox{\boldmath$v$}}\bigg\{ \ln_{_{\{{\scriptstyle
\kappa}\}}}\!\Big[\,(Zf)\oslash\mbox{\raisebox{-2.5mm}{\hspace{-3.5mm}$\scriptstyle
\kappa$}} \hspace{2mm}(Zf_s)\,\Big]\bigg\} \! \right) \ ,
\label{N201}
\end{equation}
and appears structurally similar to  Eq. (\ref{LE}).

\end{document}